\documentclass[twocolumn,prl,superscriptaddress,showpacs,amsmath,amssymb]{revtex4}
\usepackage{bm,graphicx,dcolumn}
\newcommand{\pdt}{\partial_{t}}
\newcommand{\pdx}{\partial_{x}}

\newcommand{\mC}{\bm C}
\newcommand{\mL}{\bm L}
\newcommand{\mI}{\bm I}
\newcommand{\mS}{\bm S}

\newcommand{\mN}{\bm N}
\newcommand{\vtheta}{\vec{\theta}}
\newcommand{\vpsi}{\vec{\psi}}

\newcommand{\psiop}{\hat{\psi}}

\begin{document}
\title{Variety of $c$-axis collective excitations in layered multigap superconductors}
\affiliation{
CCSE, Japan Atomic Energy Agency, 
6-9-3 Higashi-Ueno Taito-ku, Tokyo 110-0015, Japan}
\affiliation{
Institute for Materials Research, Tohoku University, 2-1-1 Katahira
Aoba-ku, Sendai 980-8577, Japan} 
\affiliation{
CREST(JST), 4-1-8 Honcho, Kawaguchi, Saitama 332-0012, Japan}
\affiliation{
JST, TRIP, 5 Sambancho Chiyoda-ku, Tokyo 102-0075, Japan}
\author{Yukihiro Ota}
\affiliation{
CCSE, Japan Atomic Energy Agency, 
6-9-3 Higashi-Ueno Taito-ku, Tokyo 110-0015, Japan}
\affiliation{
CREST(JST), 4-1-8 Honcho, Kawaguchi, Saitama 332-0012, Japan}
\author{Masahiko Machida}
\affiliation{
CCSE, Japan Atomic Energy Agency, 
6-9-3 Higashi-Ueno Taito-ku, Tokyo 110-0015, Japan}
\affiliation{
CREST(JST), 4-1-8 Honcho, Kawaguchi, Saitama 332-0012, Japan}
\affiliation{
JST, TRIP, 5 Sambancho Chiyoda-ku, Tokyo 102-0075, Japan}
\author{Tomio Koyama}
\affiliation{
Institute for Materials Research, Tohoku University, 
2-1-1 Katahira Aoba-ku, Sendai 980-8577, Japan}
\affiliation{
CREST(JST), 4-1-8 Honcho, Kawaguchi, Saitama 332-0012, Japan}
\date{\today}

\begin{abstract}
We present a dynamical theory for the phase differences along a stacked
 direction of intrinsic Josephson junctions (IJJ's) in layered multigap
 superconductors, motivated by the discovery of highly-anisotropic
 iron-based superconductors with thick perovskite-type blocking layers. 
The dynamical equations describing AC and DC intrinsic Josephson effects
 peculiar to multigap IJJ's are derived, and collective Leggett mode
 excitations in addition to the Josephson plasma established in
 single-gap IJJ's are predicted.  
The dispersion relations of their collective modes are explicitly
 displayed, and the remarkable peculiarity of the Leggett mode is demonstrated.
\end{abstract}

\pacs{74.50.+r,74.70.Xa}
\maketitle

Highly-anisotropic layered High-$T_{\rm c}$ superconductors are natural
nano-scale stacks of Josephson junctions, i.e., intrinsic Josephson
junction (IJJ) arrays, since superconducting and 
insulating layers with atomic thickness regularly alternate along the crystalline $c$-axis.  
Their high-quality single crystals clearly exhibit Josephson
effects only in $c$-axis electromagnetic response, which are called
intrinsic Josephson effects (IJE's). 
IJE's have been experimentally confirmed in various 
layered High-$T_{\rm c}$ copper oxide superconductors, such as 
$\mbox{Bi}_{2}\mbox{Sr}_{2}\mbox{CaCu}_{2}\mbox{O}_{8}$\,
\cite{Kleiner;Muller:1992,Oya;Tokutaka:1992,Tamasaku;Uchida:1992}. 
An intriguing feature in IJE's is unique dynamics arising from couplings
between the stacked junctions. 
Two types of inter-Junction couplings due to  
inductive \,\cite{Sakai;Pedersen:1993,Bulaevskii;Clem:1994} and
capacitive \,\cite{Koyama;Tachiki:1996,Machida;Tachiki:1999} origins have
been mainly proposed. 
The inductive coupling in IJJ's is very strong since the in-plane
magnetic penetration depth characterizing the magnetic field screening
range extends over several hundred junctions. 
Meanwhile, the capacitive one is not so strong since the
charge screening length is comparable to the layer thickness. 
However, it has a significant role on IJE's due to 
the atomic-scale structure\,\cite{Shukrinov;Mahfouzi:2007}.  

Most of High-$T_{\rm c}$ cuprate materials are identified as single-band
superconductor. 
One then defines just a single phase difference between consecutive
superconducting layers.  
The dynamics of the phase difference in single-gap IJJ's has been
intensively studied after the discovery of High-$T_{\rm c}$ cuprate IJJ's
\,\cite{Machida;Tachiki:2000}. 
In this paper, we extend the dynamical theory for the phase difference
to multi-gap IJJ's, in which more than one phase differences are active
through stacked all junctions. 
Our motivation comes from the discovery of
highly-anisotropic iron-based supercondcutors, such as 
$(\mbox{Fe}_{2}\mbox{As}_{2})(\mbox{Sr}_{4}\mbox{V}_{2}\mbox{O}_{6})$, 
$(\mbox{Fe}_{2}\mbox{P}_{2})(\mbox{Sr}_{4}\mbox{Sc}_{2}\mbox{O}_{6})$ 
and  $(\mbox{Fe}_{2}\mbox{As}_{2})(\mbox{Sr}_{4}\mbox{(Mg, Ti)}_{2}\mbox{O}_{6})$
\,\cite{Zhu;Wen:2009,Ogino;Shimoyama:2009,Sato;Tou:2010}.  
These compounds contain thick perovskite-type blocking layers 
$\mbox{Sr}_{4}\mbox{M}_{2}\mbox{O}_{6}$ ($\mbox{M}=\mbox{Sc, Cr, V}$)
with a thickness of $\sim 15\,\mbox{\AA}$, which clearly remind us of 
$\mbox{Bi}_{2}\mbox{Sr}_{2}\mbox{CaCu}_{2}\mbox{O}_{8}$. 
The first principles calculations on these materials indicate that they are
multiband systems with strong two-dimensional character
\,\cite{Nakamura;Hamada:2009,Nakamura;Machida:2010} whose anisotropy is 
estimated to be comparable to $\mbox{Bi}_{2}\mbox{Sr}_{2}\mbox{CaCu}_{2}\mbox{O}_{8}$.   
In fact, the experimentally observed resistivity of their
polycrystalline samples\,\cite{Zhu;Wen:2009} exhibits a
broad superconducting transition in the presence of external magnetic
field, which is a clear sign of high anisotropy. 
We also note a direct report that single crystals of an iron-based superconductor,
$\mbox{PrFeAsO}_{0.7}$, show the $I$-$V$ characteristics peculiar to
Josephson junctions in the
$c$-axis\,\cite{Kashiwaya;Kashiwaya:2010}. 

What is the most fundamental issue in multi-gap IJJ's? 
Since the tunneling channel is also multiple, the number of collective
modes in the phase oscillation is simply expected to be multiplied.  
Confining ourselves to the simplest two-gap systems, we study the
multiple collective modes. 
First, we derive coupled equations of motion for the phase
differences describing AC and DC multigap IJE's. 
Then, a mode analysis on them clarifies that the in-phase mode
corresponding to the Josephson plasma is not significantly altered while 
the out-of-phase one, i.e., Leggett mode suggested in the presence of
two superfluid orders by Leggett\,\cite{Leggett:1966}, emerges as 
a unique mode. 
An intriguing focus in this paper is that the $c$-axis Leggett mode is
weakly dispersive and favors synchronous oscillations along $c$-axis. 
Such a behavior is striking contrast to the Josephson plasma. 

\begin{figure}[tbp]
\centering
\scalebox{0.25}[0.25]{\includegraphics{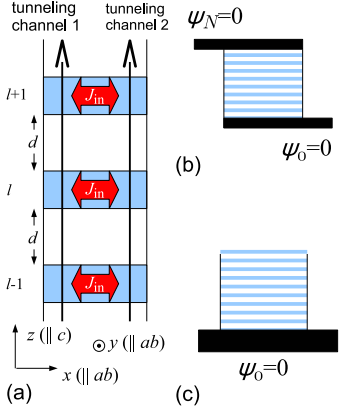}} 
\caption{(color online) (a) Schematic diagram for IJJ's with
 multiple gap superconducting layers. (b) Double-sided type junction and
 the associated boundary condition for $\psi_{l}$. (c) Mesa type
 junction and the associated boundary condition for $\psi_{l}$.}
\label{fig:ijjs} 
\end{figure}
Consider the two-gap IJJ's composed of $N$ junctions as shown in
Fig.\,\ref{fig:ijjs}(a).  
We assume the pairing interaction as
\(
-\sum_{ij=1,2}g_{ij}
\psiop^{\dagger}_{l i\uparrow }\psiop^{\dagger}_{l i\downarrow}
\psiop_{l j\downarrow}\psiop_{l j\uparrow}
\) 
on each superconducting layer, where $\psiop_{l i\sigma}$ is the
electron field operator with spin $\sigma$ in the $i$th band on the
$l$th superconducting layer. 
The coupling constants $g_{11}$ and $g_{22}$ ($g_{12}=g_{21}$) denote
the intra-band (inter-band) pairing interaction constants. 
The inter-band pairing interaction generates the inter-band Josephosn
coupling energy, 
\(
v_{\rm in}\cos(\varphi_\ell^{(1)}-\varphi_\ell^{(2)})
\), 
in the effective action of superconducting
phases\,\cite{Sharapov;Beck:2002}. 
Here, $\varphi_{l}^{(i)}$ is the phase of the superconducting gap in the
$i$th band on the $l$th superconducting layer, and the 
coupling constant $v_{\rm in}$ is given as 
\(
 v_{\rm in}=4\kappa_{\rm in}|g_{12}/g||\Delta^{(1)}||\Delta^{(2)}|
\), 
where $|\Delta^{(i)}|$ is the $i$th superconducting gap amplitude and
$g=g_{11}g_{22}-(g_{12})^{2}$. 
The coefficient $\kappa_{\rm in}$ is the sign factor defined as $\kappa_{\rm in}=1$ 
for $g_{12}>0$ and $\kappa_{\rm in}=-1$ for $g_{12}<0$.
For the Josephson coupling between consecutive superconducting layers,
one can derive the so-called Josephson coupling energy on two channels
as  
\(
 \sum_{i=1,2}(\hbar j_{c,i}/e^\ast)
\cos[\varphi_{l+1}^{(i)}-\varphi_{l}^{(i)}-(e^{\ast} d/\hbar c)A_{l+1,l}^z]
 \),
where $j_{c,1}$ and $j_{c,2}$ are the Josephson critical current
densities,  
$A_{l+1,l}^z=(1/d)\int_{l d}^{(l+1)d}A^z(z)dz$ is the $z$-component of
the vector potential, and $e^{\ast}=2e$. 
Here, we neglect the inter-band crossing channel because it is the
forth-order process in the coherent tunneling case. 
On the basis of this result, we propose an effective Lagrangian for the
two-gap IJJ's as 
\begin{eqnarray}
 L 
&=& 
\sum_{l}
\bigg[
\frac{s q_{l,1}^{2}}{8\pi \mu_{1}^{2}}
+
\frac{s q_{l,2}^{2}}{8\pi \mu_{2}^{2}}
-\frac{s v_{l,1}^{2}}{8\pi\lambda_{ab,1}^{2}}
-\frac{s v_{l,2}^{2}}{8\pi\lambda_{ab,2}^{2}}
\nonumber \\
&&
+
\frac{\hbar j_{c,1}}{e^{\ast}}
\cos\theta^{^{(1)}}_{l+1,l}
+
\frac{\hbar j_{c,2}}{e^{\ast}}
\cos\theta^{(2)}_{l+1,l}
+
\frac{\hbar J_{{\rm in}}}{e^{\ast}}
\cos\psi_{l}
 \nonumber \\
&&
+
\frac{\epsilon d}{8\pi}(E_{l+1,l}^{z})^{2} 
- 
\frac{d}{8\pi} (B_{l+1,l}^{y})^{2}
\bigg],
\label{eq:Lagrangian}
\end{eqnarray}
where 
\(
q_{l,i}
=
(\hbar/e^{\ast})\pdt\varphi_{l}^{(i)} + A^{0}_{l}
\), 
\(
v_{l,i}
=
(\hbar c/e^{\ast})\pdx\varphi_{l}^{(i)} - A^{x}_{l}
\), 
\(
\theta^{(i)}_{l+1,l}
=
\varphi_{l+1}^{(i)} - \varphi_{l}^{(i)}
-
(e^{\ast}d/\hbar c) A^{z}_{l+1,l}
\), 
\(
J_{\rm in} = e^{\ast}v_{\rm in}s/\hbar
\),
and 
\(
\psi_{l}=\varphi^{(1)}_{l} - \varphi^{(2)}_{l}
\). 
The parameters, $s$ and $d$, are the thicknesses of the superconducting
and insulating layers, respectively, $\mu_{i}$ ($\lambda_{ab,i}$) is the
charge screening length (in-plane penetration depth) relevant to the
$i$th-band electrons, $\epsilon$ is the dielectric constant in the insulating layers, and 
$A^{0}_{l}$ and $A^{x}_{l}$ are, respectively, the scalar potential and  the
$x$-component of the vector potential on the $l$th superconducting layer.
Without losing generality, we consider only the $z$-component of the
electric field and only the $y$-component of the magnetic field, which
are expressed as  
\(
E_{l+1,l}^{z}
=
-(1/c)\pdt A^{z}_{l+1,l} - (A^{0}_{l+1}-A^{0}_{l})/d
\) and 
\(
B_{l+1,l}^{y}
=
(A^{x}_{l+1}-A^{x}_{l})/d - \pdx A^{z}_{l+1,l}
\). 
As in the single-gap IJJ's one can define the
inductive\,\cite{Sakai;Pedersen:1993,Bulaevskii;Clem:1994} 
and capacitive\,\cite{Koyama;Tachiki:1996,Machida;Tachiki:1999} coupling
constants in the dimensionless form as  
\(
\eta_{i} = \lambda_{ab,i}^{2}/sd
\) 
and 
\(
\alpha_{i} = \epsilon \mu_{i}^{2}/sd
\) for each channel in this system. 
The effective action (\ref{eq:Lagrangian}) describes 
low energy dynamics of the superconducting phases in the two-gap IJJ's. 
In the derivation of Eq.\,(\ref{eq:Lagrangian}), all junction parameters
(e.g. $j_{{\rm c},i}$ and $\lambda_{{\rm ab},i}$) are approximated as
local quantities for brevity, although they are originally nonlocal
ones. 
As for the nonlocal electromagnetic effects, see
Ref.\,\cite{Abdumalikov;Malishevskii:2009}. 

Now, let us derive the coupled equations of motion of the
superconducting phase differences. 
First, one obtains the so-called Josephson relations associated with time and 
spatial variations of the superconducting phase differences as, 
\begin{subequations}
\label{eq:gr}
\begin{eqnarray}
&&
\pdt \theta_{l+1,l} - \frac{\xi}{2}\pdt\psi_{l+1,l}
=
\frac{e^{\ast}d}{\hbar}
(1-\tilde{\alpha}\triangle^{(2)})E^{z}_{l+1,l},
\label{eq:gr_e}\\
&&
\pdx \theta_{l+1,l} - \frac{\zeta}{2}\pdt\psi_{l+1,l}
=
\frac{2\pi d}{\Phi_{0}}
(1-\tilde{\eta}\triangle^{(2)}) B^{y}_{l+1,l},
\label{eq:gr_m}
\end{eqnarray} 
\end{subequations} 
where $\tilde{\alpha}$ and $\tilde{\eta}$ are, respectively, the reduced
capacitive and inductive coupling constants given as 
\(
\tilde{\alpha}^{-1} = \alpha_{1}^{-1} + \alpha_{2}^{-1}
\)
and 
\(
\tilde{\eta}^{-1} = \eta_{1}^{-1} + \eta_{2}^{-1}
\), 
\(
\triangle^{(2)}
\) the second-order finite difference defined as 
\(
\triangle^{(2)}f_{l+1,l}
=f_{l+2,l+1} - 2f_{l+1,l} + f_{l,l-1}
\) (for $^{\forall}f_{l+1,l}$),  $\Phi_{0}(=2\pi\hbar c/e^{\ast})$ the
unit flux, and  
\(
\xi=(\alpha_{1}-\alpha_{2})/(\alpha_{1}+\alpha_{2})
\)
and 
\(
\zeta=(\eta_{1}-\eta_{2})/(\eta_{1}-\eta_{2})
\). 
Here, we introduce 
\(
\theta_{l+1,l} = (\theta^{(1)}_{l+1,l}+\theta^{(2)}_{l+1,l})/2
\) 
and 
\(
\psi_{l+1,l} = \psi_{l+1} - \psi_{l}
\). 
Equations (\ref{eq:gr_e}) and (\ref{eq:gr_m}) are interpreted as the
generalized Josephson relations in the two-gap IJJ's. These relations
are reduced to the conventional ones in the single-gap IJJ's when $\xi=\zeta=0$. 
On the variation of $A_{l+1,l}^z$, the Lagrangian (\ref{eq:Lagrangian})
derives the Maxwell equation as 
\begin{eqnarray}
\pdx B^{y}_{l+1,l} 
-
\frac{\epsilon}{c}\pdt E^{z}_{l+1,l}
=
\frac{4\pi}{c}(j^{\rm J}_{l+1,l} + j^{\rm QP}_{l+1,l}),
\label{eq:Maxwell}
\end{eqnarray}
where 
\(
j^{\rm J}_{l+1,l}
=
\sum_{i=1}^{2} j_{c,i}\sin\theta^{(i)}_{l+1,l}
\) 
and 
\(
j^{{\rm QP}}_{l+1,l}
=
\sum_{i=1}^{2} j^{{\rm QP}(i)}_{l+1,l}
\). 
Here, we add the quasi-particle tunneling current 
$j_{l+1,l}^{\rm QP}$, which can be derived microscopically\,\cite{Machida;Tachiki:2000}. 
Furthermore, we have the continuity equations, which can be derived by
the variation with respect to $\varphi_{l}^{(i)}$\,\cite{Machida;Tachiki:2000}. 
From the continuity equations with Eq.\,(\ref{eq:gr}) we also have the
``pseudo'' Maxwell equation, which describe the motion of the relative
phase differences $\psi_{l+1,l}$, as 
\begin{eqnarray}
&&
\pdx \widetilde{B}^{y}_{l+1,l}
-
\frac{\epsilon}{c}\pdt \widetilde{E}^{z}_{l+1,l}
=
\frac{4\pi}{c}2J_{\rm in}
(
\sin\psi_{l+1}-\sin\psi_{l}
) 
\nonumber \\
&&
\hspace{33mm}
+
\frac{4\pi}{c}\triangle^{(2)}
(d^{\rm J}_{l+1,l} + d^{\rm QP}_{l+1,l}), 
\label{eq:rlt_diff_eq} 
\end{eqnarray} 
where
\(
d^{\rm J}_{l+1,l}
=
-j_{c,1}\sin\theta^{(1)}_{l+1,l}
+j_{c,2}\sin\theta^{(2)}_{l+1,l}
\), 
\(
d^{\rm QP}_{l+1,l}
=
-j^{{\rm QP}(1)}_{l+1,l}
+j^{{\rm QP}(2)}_{l+1,l}
\).
The ``pseudo'' electromagnetic fields $\widetilde{E}^{z}_{l+1,l}$ and
$\widetilde{B}^{y}_{l+1,l}$ are defined as 
\begin{eqnarray*}
&&
\widetilde{E}^{z}_{l+1,l}
= 
\frac{\hbar}{e^{\ast}}
\frac{1-\xi^{2}}{\tilde{\alpha}d}
\pdt \psi_{l+1,l}
+
\xi \triangle^{(2)}E^{z}_{l+1,l},\\
&&
\widetilde{B}^{y}_{l+1,l}
=
\frac{\Phi_{0}}{2\pi}
\frac{1-\zeta^{2}}{\tilde{\eta}d}
\pdx \psi_{l+1,l}
+
\zeta \triangle^{(2)} B^{y}_{l+1,l}. 
\end{eqnarray*}
Equations (\ref{eq:gr}), (\ref{eq:Maxwell}), and (\ref{eq:rlt_diff_eq})
provide a set of equations of motion for the phase differences and the
electromagnetic fields in the two-gap IJJ's, that is, the DC and AC
Josephson effects in the two-gap IJJ's can be described by these coupled
equations. 
To solve these equations it is convenient to use the relation defined as 
\(
 \psi_{l}= \sum_{m=1}^{l}\psi_{m,m-1} + \psi_{0}
\), where the value of $\psi_0$ is specified
as the boundary condition [Figs.\,\ref{fig:ijjs}(b) and
\ref{fig:ijjs}(c)].  

\begin{figure}[bp]
\begin{tabular}{cc}
(a)\!\!\!\scalebox{0.70}[0.70]{\includegraphics{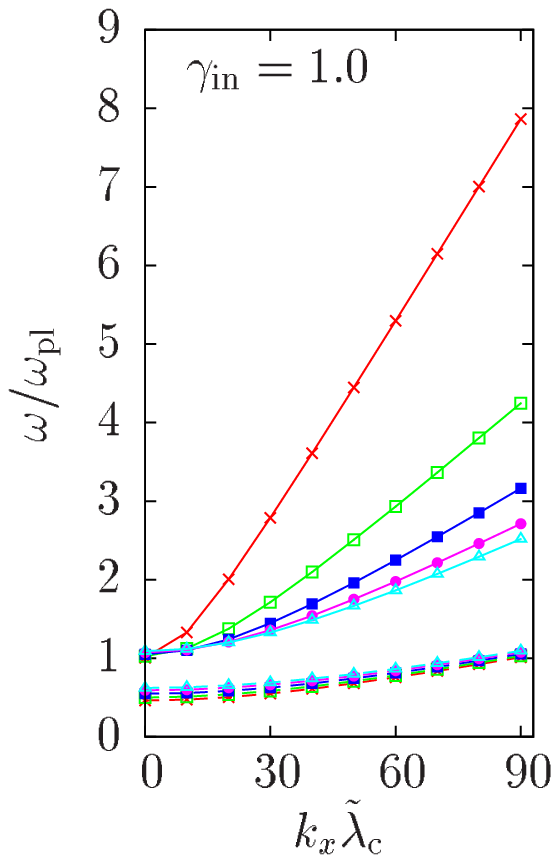}} & 
\begin{tabular}[b]{c}
(b)\!\!\!\scalebox{0.48}[0.48]{\includegraphics{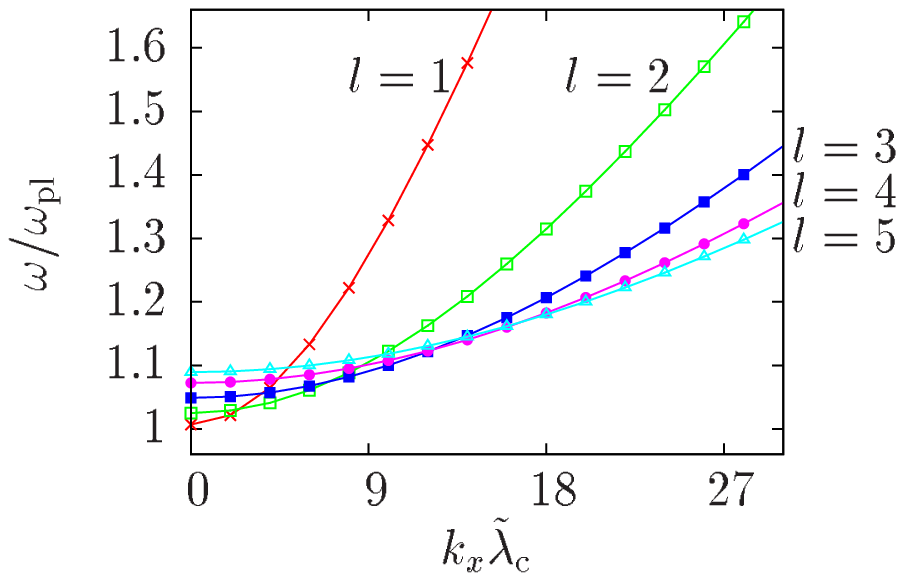}} \\
\\
(c)\!\!\!\scalebox{0.48}[0.48]{\includegraphics{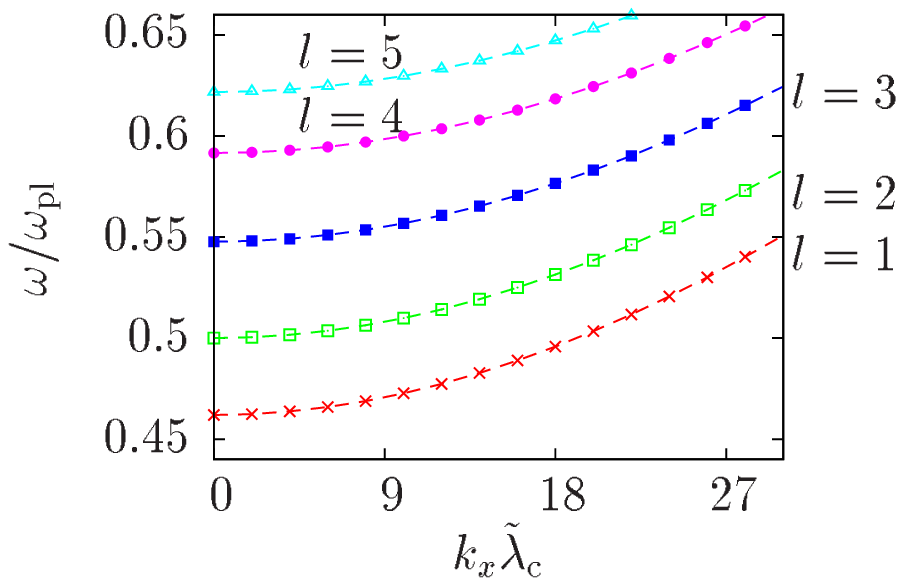}}
\end{tabular}  
\end{tabular} 
\caption{(color online) (a) Dispersion relations for the Josephson-plasma (solid lines)
 and the Leggett's modes (dash lines) when $N=5$ and 
$\gamma_{\rm in}=1.0$. We set $\alpha_{1}=\alpha_{2}=0.1$,
 $\eta_{1}=\eta_{2}=10^{3}$, and $j_{c,1}=j_{c,2}$. 
 Enlarged views of the dispersion relations at small in-plane wavenumbers 
 for  the five eigenmodes of the Josephson-plasma (b) and the  Leggett's modes (c).  }
\label{fig:disp1}
\end{figure} 

Let us focus on the collective phase oscillation modes in the two-gap IJJ's. 
Consider the $N$ junction system under the periodic boundary condition
along the $c$-axis. 
For simplicity, the case of $\xi=\zeta=0$ and $j_{c,1}=j_{c,2}$ is
examined in the following. 
More general cases will be published elsewhere. 
Assuming small oscillations, 
we linearize Eqs.\,(\ref{eq:Maxwell}) and
(\ref{eq:rlt_diff_eq}) around $\theta_{l+1,l}=0$ and $\psi_{l}=0$ 
with neglecting the dissipation currents $j_{\ell+1,\ell}^{\rm QP}$ and
$d_{l+1,l}^{\rm QP}$ for the standard mode analysis\,\cite{Fetter;Stephen:1968}. 
The dynamical simulation taking into account the quasiparticle contributions
was performed in Ref.\,\cite{Koyama;Machida:2010}.  
Eliminating the electric and magnetic fields from the coupled
linearized equations, we can derive the decoupled equations for
$\theta_{l+1,l}$ and $\psi_{l+1,l}$ as follows, 
\begin{subequations}
\label{eq:l_eq}
\begin{eqnarray}
&&
\mC \pdx^{2}\vtheta - \frac{\epsilon}{c^{2}}\mL \pdt^{2}\vtheta
=
\frac{1}{\tilde{\lambda}_{c}^{2}}\mC\mL \vtheta, 
\label{eq:l_Maxwell}
\\
&&
\frac{1}{2\tilde{\eta}}\mI \pdx^{2}\vpsi
-
\frac{1}{2\tilde{\alpha}}\mI \pdt^{2}\vpsi
=
\frac{2}{\lambda_{\rm in}^{2}}
\mN \vpsi,
\label{eq:l_rlt_diff_eq}
\end{eqnarray}
\end{subequations} 
where 
\(
\tilde{\lambda}_{c}^{-2} = \lambda_{c,1}^{-2} + \lambda_{c,2}^{-2}
\), 
\(
\lambda_{c,i}^{-2}
=4\pi e^{\ast}dj_{c,i}/\hbar c^{2}
\), 
\(
\lambda_{\rm in}^{-2}
=4\pi e^{\ast}d |J_{\rm in}|/\hbar c^{2}
\), 
\(
\vtheta = \!^{t}(\theta_{2,1},\theta_{3,2},\ldots,\theta_{N-1,N})
\), and 
\(
\vpsi = \!^{t}(\psi_{2,1},\psi_{3,2},\ldots,\psi_{N-1,N})
\). 
The coeffcients, $\mC$ and $\mL$, are $N\times N$ matrices given as 
\(
\mC = (1+2\tilde{\alpha})\mI - \tilde{\alpha}\mS
\) and 
\(
\mL = (1+2\tilde{\eta})\mI - \tilde{\eta}\mS
\), 
where $\mS$ is an $N\times N$ tridiagonal matrix with the elements,
$S_{l,l}=0$ and $S_{l,l\pm 1}=1$, and $\mI$ is the $N\times N$ unit
matrix. 
We note that the matrices $\mC$ and $\mL$ represent, respectively, the
capacitive and inductive couplings between junctions, which are the same
as those in the single-gap IJJ's.  
Thus, the collective motion of the mean phase differences, which is described by 
Eq.\,(\ref{eq:l_Maxwell}), is understood to be the Josephson plasma. 
Moreover, it is clearly found that its dispersion is brought about by
the inductive and capacitive couplings between junctions.  
On the other hand, due to two-gap IJJ's, we have another collective
oscillation mode in the relative phase channel, which is described by
Eq.\,(\ref{eq:l_rlt_diff_eq}). 
In the new mode, its origin, i.e., the coupling between junctions is
found to be induced by the off-diagonal components of the matrix 
\(
\mN = (1 + 2\nu) \mI - \nu \mS
\) with 
\begin{equation}
\nu = \frac{1}{4\gamma_{\rm in}^{2}}, 
\quad
\gamma_{\rm in} 
= \frac{\tilde{\lambda}_{c}}{\lambda_{\rm in}}
= \sqrt{\frac{|J_{\rm in}|}{j_{c,1}+j_{c,2}}}. 
\label{eq:def_gamma_in}
\end{equation}
We note that the coupling constant $\gamma_{\rm in}$ (or $\nu$) depends
on not the inductive and capacitive coupling constants but just the
inter-band Josephson coupling $J_{\rm in}$. 
Thus, this mode has its origin only in the
inter-band pairing interaction.  
Hence, one understands that Eq.\,(\ref{eq:l_rlt_diff_eq}) describes the
Leggett mode in the two-gap IJJ's. 
From these results, it is concluded that the Josephson plasma mode is
originated from the inductive and capacitive coupling arising from the
electromagnetic field screening, while the Leggett mode is brought
about by the intra-layer interband coupling.  

\begin{figure}[tbp]
\begin{tabular}{cc}
(a)\!\!\!\scalebox{0.70}[0.70]{\includegraphics{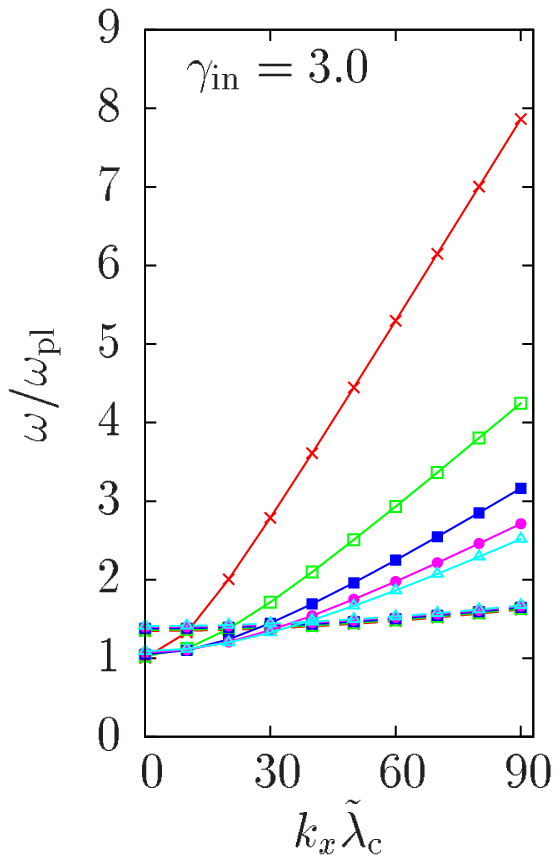}} & 
\begin{tabular}[b]{c}
(b)\!\!\!\scalebox{0.48}[0.48]{\includegraphics{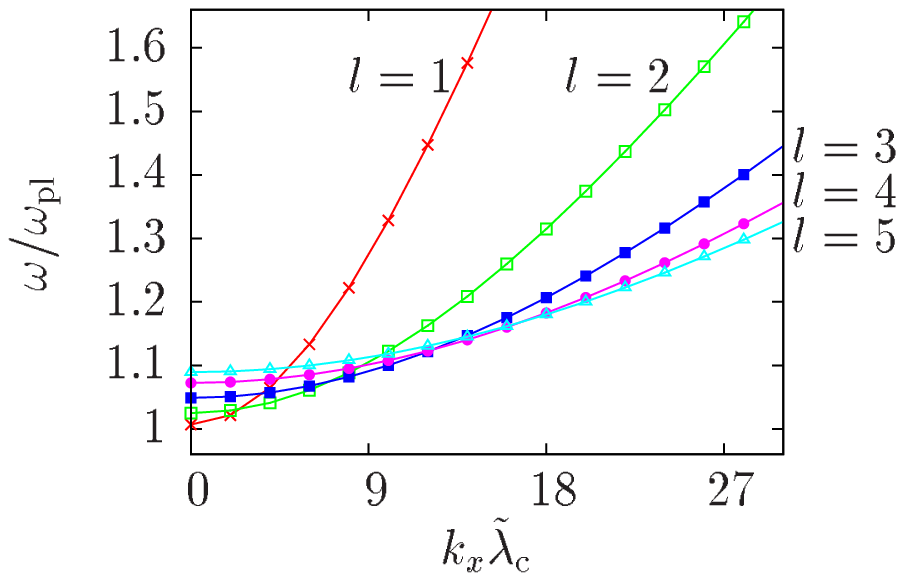}} \\
\\
(c)\!\!\!\scalebox{0.48}[0.48]{\includegraphics{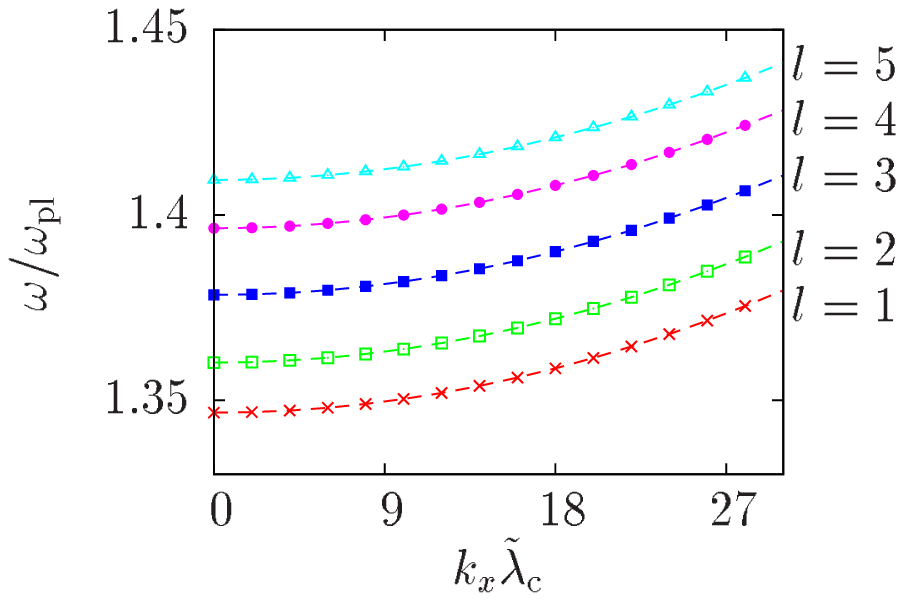}}
\end{tabular}  
\end{tabular} 
\caption{(color online) (a) Dispersion relations for the Josephson-plasma (solid lines)
 and the Leggett's modes (dash lines) when $\gamma_{\rm in}=3.0$. 
The other parameter values are the same as in  Fig.\,\ref{fig:disp1}. 
Enlarged viewe at small wave numbers are shown in (b)  and (c).}
\label{fig:disp3}
\end{figure}

The dispersion relations of these two eigen-modes are obtained
from  Eqs.\,(\ref{eq:l_Maxwell}) and (\ref{eq:l_rlt_diff_eq}), which are
specified in terms of the wave numbers $k_x$ (in-plane direction) and
$k_z=l\pi/(N+1)d$ ($c$-direction) as  
\begin{equation}
\omega_{\rm P}(k_{x},l)
=
\omega_{\rm P}(0,l)
\sqrt{
1 + \frac{k_{x}^{2} \tilde{\lambda}_{c}^{2}}{
1+2\tilde{\eta}(1-s_{l})}},
\label{eq:disp_plasma}
\end{equation} 
with 
\(
\omega_{\rm P}(0,l)
=
\omega_{\rm pl}
\sqrt{
1+2\tilde{\alpha}(1-s_{l})}
\) for the longitudinal Josephson plasma and 
\begin{equation}
\omega_{\rm L}(k_{x},l)
=
\omega_{\rm L}(0,l)
\sqrt{
1
+
\frac{k_{x}^{2}\lambda_{\rm Leg}^{2}}
{1+2\nu (1-s_{l})}}, 
\label{eq:disp_leggett}
\end{equation} 
with 
\(
\lambda_{\rm Leg}
=
2\sqrt{\nu}\tilde{\lambda}_{c}/\sqrt{\eta_{1}+\eta_{2}}
\) and 
\(
\omega_{\rm L}(0,l)
=
\omega_{\rm Leg}\sqrt{
1+
2\nu (1-s_{l})}
\) for the longitudinal Leggett mode, 
where  
\(
s_{l} = \cos[l\pi/(N+1)]
\) 
and $\omega_{\rm pl}$ and $\omega_{\rm Leg}$ are, 
respectively, the Josephson plasma and the Leggett mode frequencies,
i.e., 
\(
\omega_{\rm pl}=c/\sqrt{\epsilon}\tilde{\lambda}_{c}
\)
and
\(
 \omega_{\rm Leg} 
=
c\sqrt{\alpha_{1}+\alpha_{2}}/\sqrt{\epsilon}\lambda_{\rm in} 
\). 
Here, it is clearly found that the origin of the Leggett mode is a fluctuation between 
two superfluids which is essential to 
neutral multiple superfluids.

We plot the dispersion relations of these eigen modes in the case of $N=5$ with 
\(
j_{c,1}=j_{c,2}
\)
for $\gamma_{\rm in}=1.0$ and $3.0$, respectively, in
Figs.\,\ref{fig:disp1} and \ref{fig:disp3}.  
The values of the inductive and capacitive coupling constants are chosen as  
\(
\alpha_{1}=\alpha_{2}=0.1
\) and
\(
\eta_{1}=\eta_{2}=10^{3}
\), 
which are the values applicable to the cuprate IJJ's. If the Leggett
mode is a low-energy excitation mode and can lie inside the energy gaps as the Josephon
plasma, then it is possible that both modes are closely located in
the low energy range as shown in these figures.  
It should be also noted that the Josephson plasma mode with the largest $c$-axis 
wave number, i.e., $l=5$, is the lowest energy one close to $k_x=0$,
but this mode changes to the highest one for larger values of $k_x$. 
This is because the large inductive coupling, which is predominant in a
wide $k_{x}$ range, favors $\pi$ phase shift in the
phase differences between consecutive junctions\,\cite{Machida;Sakai:2004}. 
This discussion clearly leads to that $\pi$ anit-phase synchronization is preferable
in the Josephson plasma mode under the presence of the layer parallel magnetic field.
In fact, strong synchronous electromagnetic-wave emission is observed
only at the zero and weak field in layered High-$T_{\rm c}$ copper
oxide superconductors\,\cite{Ozyuzer;Welp:2007}. 
On the other hand, the dispersions in the Leggett mode does not 
show such level crossing as seen in the figures, since the excitation
mode is associated with only the density channel. 
This indicates that the Leggett excitation always prefers synchronous oscillations along
junction stacked direction even in the presence of the magnetic field.
If the Leggett mode is excited by the charge injection or other ways,
then the synchronized Leggett oscillation emerges and a conversion into
the synchronized Josephson plasma excitation due to inherent
nonlinearity may occur. 

Finally, we mention that when the difference between the two tunneling
channels exist (i.e., $j_{{\rm c},1}\neq j_{{\rm c},2}$, 
$\alpha_{1}\neq \alpha_{2}$, and $\eta_{1}\neq \eta_{2}$) a mode
coupling between the Josephson-plasma and the Leggett modes can occur. 
Such a coupling effect is an interesting future task.

In summary, we derived the coupled dynamical equations for 
the phase differences which can be utilized for the analysis of AC and
DC Josephson effects in the multi-gap IJJ's. The equations revealed that
multi-gap IJJ's have two collective phase oscillation modes, the
Josephson plasma and the Leggett mode whose origins are different. 
Moreover, it is found the Josephson plasma and Leggett modes favor $\pi$
anti-phase and in-phase synchronization along the junction stacking ,
respectively, in a wide wave-number range. 

YO and MM would like to thank Hiroki Nakamura for helpful discussions. 
TK was partially supported by Grant-in-Aid for Scientific Research (C)
(No. 22540358) from the Japan Society for the Promotion of Science.

\end{document}